\documentclass[12pt,preprint]{aastex}
\usepackage{longtable}

\def\gtorder{\mathrel{\raise.3ex\hbox{$>$}\mkern-14mu
             \lower0.6ex\hbox{$\sim$}}} 
\def\ltsima{$\; \buildrel < \over \sim \;$}
\def\simlt{\lower.5ex\hbox{\ltsima}}
\def\gtsima{$\; \buildrel > \over \sim \;$}
\def\simgt{\lower.5ex\hbox{\gtsima}} 

\begin{document} 


\title{PTF11eon/SN2011dh: Discovery of a Type IIb Supernova from a Compact Progenitor in the Nearby Galaxy M51}


\author{Iair Arcavi, Avishay Gal-Yam, Ofer Yaron, Assaf Sternberg, Itay Rabinak, Eli Waxman}
\affil{Department of Particle Physics and Astrophysics, The Weizmann Institute of Science, Rehovot 76100, Israel}
\email{iair.arcavi@weizmann.ac.il}

\author{Mansi M. Kasliwal, Robert M. Quimby}
\affil{Cahill Center for Astrophysics, California Institute of Technology, Pasadena, CA, 91125, USA}

\author{Eran O. Ofek}
\affil{Cahill Center for Astrophysics, California Institute of Technology, Pasadena, CA, 91125, USA}
\affil{Einstein Fellow}

\author{Assaf Horesh, Shrinivas R. Kulkarni}
\affil{Cahill Center for Astrophysics, California Institute of Technology, Pasadena, CA, 91125, USA}

\author{Alexei V. Filippenko, Jeffrey M. Silverman, S. Bradley Cenko, Weidong Li, Joshua S. Bloom}
\affil{Department of Astronomy, University of California, Berkeley, CA 94720-3411, USA}

\author{Mark Sullivan}
\affil{Department of Physics (Astrophysics), University of Oxford, Keble Road, Oxford, OX1 3RH, UK}

\author{Derek B. Fox}
\affil{Astronomy and Astrophysics, Eberly College of Science, The Pennsylvania State University, University Park, PA 16802, USA}

\author{Peter E. Nugent}
\affil{Department of Astronomy, University of California, Berkeley, CA 94720-3411, USA}
\affil{Computational Cosmology Center, Lawrence Berkeley National Laboratory, 1 Cyclotron Road, Berkeley, CA 94720, USA}

\author{Dovi Poznanski}
\affil{Department of Astronomy, University of California, Berkeley, CA 94720-3411, USA}
\affil{Computational Cosmology Center, Lawrence Berkeley National Laboratory, 1 Cyclotron Road, Berkeley, CA 94720, USA}
\affil{Einstein Fellow}

\author{Evgeny Gorbikov}
\affil{The Wise Observatory and the Raymond and Beverly Sackler School of Physics and Astronomy, the Faculty of Exact Sciences, Tel Aviv University, Tel Aviv 69978, Israel}

\author{Amedee Riou}
\affil{Asnieres 49370, Becon les Granits, France}

\author{Stephane Lamotte-Bailey}
\affil{14 rue de Dammarie, 77000 Melun, France}

\author{Thomas Griga}
\affil{Am Derkmannsstueck 34, Schwerte, D-58239, Germany}

\author{Judith G. Cohen}
\affil{Cahill Center for Astrophysics, California Institute of Technology, Pasadena, CA, 91125, USA}

\author{David Polishook, Dong Xu, Sagi Ben-Ami, Ilan Manulis}
\affil{Department of Particle Physics and Astrophysics, The Weizmann Institute of Science, Rehovot 76100, Israel}

\author{Emma S. Walker}
\affil{Scuola Normale Superiore di Pisa, Pisa 56126, Italy}

\author{Paolo A. Mazzali}
\affil{INAF, Osservatorio Astronomico di Padova, Italy}
\affil{Max-Planck Institute for Astrophysics, Garching, Germany}

\author{Elena Pian}
\affil{Scuola Normale Superiore di Pisa, Pisa 56126, Italy}
\affil{INAF, Astronomical Observatory of Trieste, Via G.B. Tiepolo 11, I-34143 Trieste, Italy}

\author{Thomas Matheson}
\affil{National Optical Astronomy Observatory, NOAO System Science Center, 950 North Cherry Avenue, Tucson, AZ 85719, USA}

\author{Kate Maquire, Yen-Chen Pan}
\affil{Department of Physics (Astrophysics), University of Oxford, Keble Road, Oxford, OX1 3RH, UK}

\author{David Bersier, Philip James, Jonathan M. Marchant, Robert J. Smith, Chris J. Mottram, Robert M. Barnsley}
\affil{Astrophysics Research Institute, Liverpool John Moores University, Twelve Quays House, Egerton Wharf, Birkenhead, CH41 1LD, UK}

\and

\author{Michael T. Kandrashoff, Kelsey I. Clubb}
\affil{Department of Astronomy, University of California, Berkeley, CA 94720-3411, USA}



\newpage

\begin{abstract} 

On May 31, 2011 UT a supernova (SN) exploded in the nearby galaxy M51 (the Whirlpool Galaxy). We discovered this event using small telescopes equipped with CCD cameras, as well as by the Palomar Transient Factory (PTF) survey, and rapidly confirmed it to be a Type II supernova. Our early light curve and spectroscopy indicates that PTF11eon resulted from the explosion of a relatively compact progenitor star as evidenced by the rapid shock-breakout cooling seen in the light curve, the relatively low temperature in early-time spectra, and the prompt appearance of low-ionization spectral features. The spectra of PTF11eon are dominated by H lines out to day 10 after explosion, but initial signs of He appear to be present. Assuming that He lines continue to develop in the near future, this SN is likely a member of the cIIb (compact IIb; Chevalier and Soderberg 2010) class, with progenitor radius larger than that of SN 2008ax and smaller than the eIIb (extended IIb) SN 1993J progenitor. Our data imply that the object identified in pre-explosion {\it Hubble Space Telescope} images at the SN location is possibly a companion to the progenitor or a blended source, and not the progenitor star itself, as its radius ($\sim10^{13}$\,cm) would be highly inconsistent with constraints from our post-explosion photometric and spectroscopic data.

\end{abstract} 




\section{Introduction} 

At the end of their lives, massive stars ($M \gtrsim 8\, {\rm M}_\odot$) explode as core-collapse supernovae (SNe), leaving a neutron star or a black hole remnant after the explosion. Several different types of SNe have been observed, including H-rich Type II-P/II-L SNe, H-rich interacting Type IIn SNe, H-poor Type IIb SNe, H-less Type Ib SNe and H- and He-less Type Ic SNe (see Filippenko 1997 for a review). 

Direct progenitor detections in pre-explosion images have established a link between low-mass red supergiants (RSGs) and Type IIP SNe, the most common type of core-collapse SN (see Smartt 2009 for a review). Evidence also exists linking Type IIn SNe with much more massive progenitors (Gal-Yam et al. 2007; Gal-Yam \& Leonard 2009; Smith et al. 2011a). However, the mapping between massive star classes and types of SNe is far from complete, due to both theoretical gaps and to the scarcity of observational constraints. Numerous core-collapse SNe are discovered each year and statistical analyses of large samples are yielding interesting insights (e.g., Anderson \& James 2009; Boissier \& Prantzos 2009; Arcavi et al. 2010; Li et al. 2011; Smith et al. 2011b). However, some of the most robust information comes from studying the few nearest events.

The Palomar Transient Factory (PTF; Rau et al. 2009; Law et al. 2009) is a wide-field variability survey. Our short cadence strategy and real-time capability (Gal-Yam et al. 2011) enables the discovery of SNe at early stages of the explosion. Some parts of the sky, such as those with photogenic nearby galaxies, are observed by amateur astronomers even more frequently. Thus, the amateur community may be the first to discover and report bright events in these nearby hosts. Indeed, this is the case here.

We report the discovery of PTF11eon / SN2011dh (CBET 2736; Silverman et al. 2011), a Type IIb supernova in M51, a very nearby interacting spiral galaxy located at a distance of $8.03\pm 0.77$ Mpc (retrieved from the NASA/IPAC Extragalactic Database\footnote{NED; http://nedwww.ipac.caltech.edu/.}). We find rapid light curve evolution at early times and relatively cool early spectra. Our data suggest that PTF11eon resulted from a progenitor much more compact than that of the prototypical Type IIb SN 1993J, but perhaps more extended than the progenitor of the Type IIb SN 2008ax.

This discovery should facilitate numerous follow-up studies. In a companion paper we report the possible progenitor system of this event, detected in archival {\it Hubble Space Telescope (HST)} imaging (Van Dyk et al. 2011).

\section{Discovery}

\begin{figure}
\includegraphics[width=165mm]{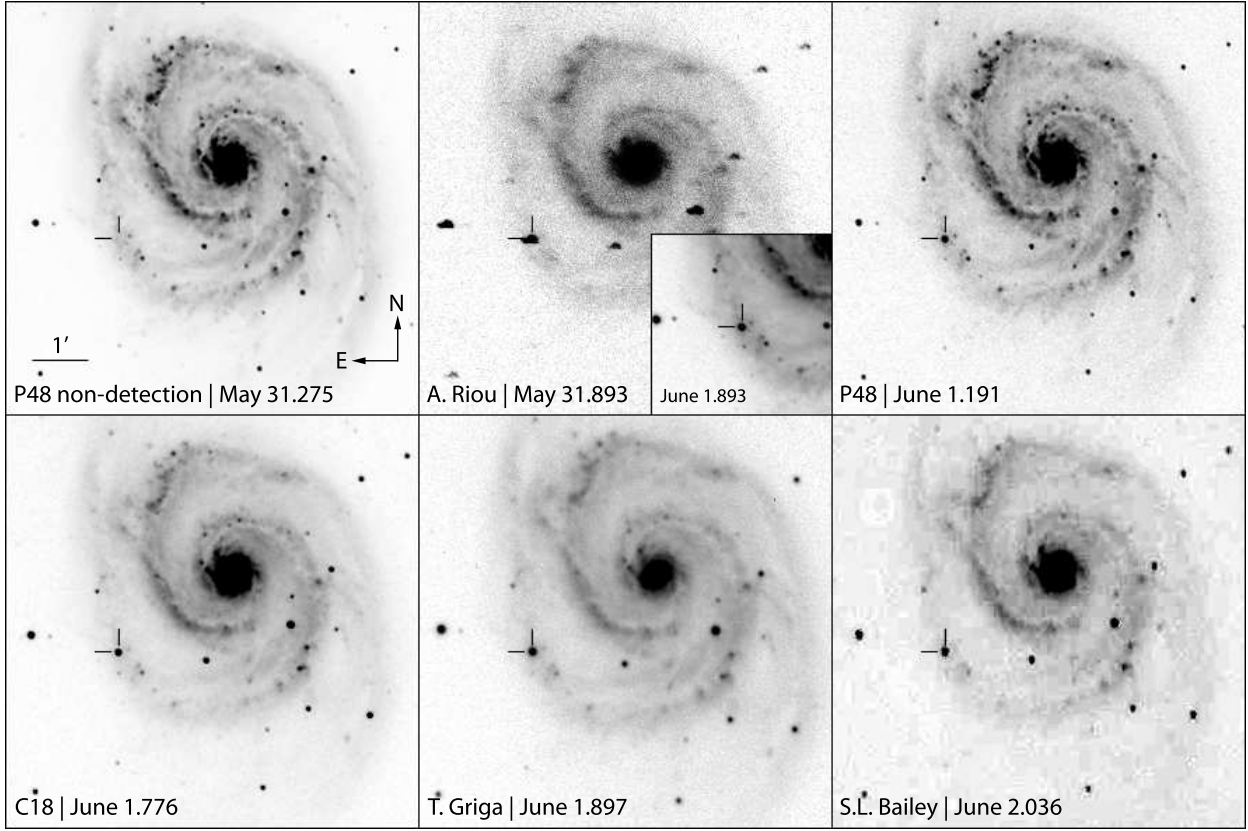}
\caption{Discovery of PTF11eon by both professional and amateur astronomers. The first reported discovery image was taken by A. Riou less than 15 hours after the PTF non-detection (note that this image was obtained while testing new equipment without guiding; a high quality confirmation image taken by A. Riou is shown in the inset). Subsequent discovery and confirmation images are also shown. UT dates are used throughout.}
\label{discovery}
\end{figure}

The supernova was discovered in images taken on May 31.893 (UT times are used throughout this paper) by A. Riou in France\footnote{http://www.cbat.eps.harvard.edu/unconf/followups/J13303600+4706330.html} (Fig. \ref{discovery}). Approximately seven hours later, on 2011 June 1.191, the PTF detected the supernova in images taken by the Palomar 48\,in Oschin Schmidt Telescope (P48). The SN was seen at $\alpha$(J2000) = $13^{\rm h}30^{\rm m}05.08^{\rm s}$ and $\delta$(J2000) = $+47^\circ10'11.2''$ at a magnitude of $m_g=13.15\pm0.09$, and it was absent from a PTF image taken by the P48 on 2011 May 31.275 down to a limiting magnitude of $m_g=21.44$. Subsequent imaging was undertaken at the Wise Observatory 18\,in telescope on June 1.776 and by T. Griga on June 1.897 and S. L. Bailey on June 2.036. Data collected by amateur astronomers during the first hours of this event are very helpful in determining the explosion time. The data described above constrain the explosion time to be between May 31.275 and May 31.893; however, this time window can still be decreased (Gal-Yam et al. in prep.).
PTF11eon is the third supernova to be discovered in M51 in the last 17 years, after the sub-luminous Type IIP SN 2005cs and the Type Ic SN 1994I.

Li \& Filippenko (2011a, 2011b) identify a star at the approximate location of the SN in pre-explosion {\it HST} images. Maund et al. (2011) confirm this with adaptive optics to within $23$\,mas and Van Dyk et al. (2011) improve the location accuracy to $7$\,mas. Maund et al. (2011) further claim that this star, a F8 supergiant with ${\log} (L/{L}_{\odot}) = 4.92 \pm 0.20$ and $T_{\rm eff}=6000\pm280$\,K (which implies a radius of $\sim10^{13}$\,cm), is the progenitor of PTF11eon. We show that this is unlikely.

\section{Follow-Up Observations}

Immediately after discovery, we initiated an extensive follow-up campaign. A spectrum was obtained with the Low Resolution Imaging Spectrometer (LRIS; Oke et al. 1995) spectrograph mounted on the Keck-I 10\,m telescope on June 3.304, identifying the event as a Type II supernova (Silverman et al. 2011). Additional spectra were later obtained with the Kast spectrograph (Miller \& Stone 1993) on the Lick Observatory 3\,m Shane telescope, the DOLORES spectrograph on the TNG 3.6\,m telescope, the RC spectrograph on the KPNO 4\,m Mayall telescope, the ISIS spectrograph on the William Herschel 4.2\,m telescope and the FRODOspec spectrograph (Morales-Rueda et al. 2004) on the Liverpool 2\,m Telescope (LT). All spectral frames were wavelength calibrated using standard arc-lamp spectra. Flux calibrations were applied using a solution derived from standard stars. All of the reductions were carried out using standard IRAF and IDL routines.

In order to estimate the extinction in the vicinity of the SN, we obtained two high-resolution spectra using the High Resolution Echelle Spectrometer (HIRES; Vogt et al. 1994) mounted on the Keck-I 10\,m telescope. The spectral resolution was $R=\lambda/\delta\lambda = 36000$ ($\delta\lambda=0.164$\,\AA\, in the vicinity of the Na~I~D $\lambda\lambda5890,5896$\,\AA\, lines). The data were reduced using the MAuna Kea Echelle Extraction (MAKEE) pipeline, written by T. Barlow\footnote{http://spider.ipac.caltech.edu/staff/tab/makee/}. The spectra were flux normalized and the Na~I~D absorption equivalent width was measured using the IRAF SPLOT tool. 

Optical and infrared photometry was obtained at P48, the Wise Observatory 1\,m (using LAIWO; Baumeister et al. 2006; Gorbikov et al. 2010) and 18\,in (C18; Brosch et al. 2008) telescopes and the Peters Automated InfraRed TELescope (PAIRITEL; Bloom et al. 2006). Our photometry is performed on difference images by constructing a reference frame from data taken before the SN explosion, and then subtracting that from each frame in which the SN is present, matching the point-spread functions (PSFs) of the two images (Gal-Yam et al. 2004; Gal-Yam et al. 2008). For the P48 data, the SN photometry was measured using a PSF fitting method. In each image frame, the PSF was determined from nearby field stars, and this average PSF was then fit at the position of the SN, weighting each pixel according to Poisson statistics. Saturated pixels (present only in the P48 data) have their weight set to zero. We calibrate our P48 light curve to the Sloan Digital Sky Survey (SDSS\footnote{http://www.sdss.org.}; York et al. 2000) system using SDSS observations of the same field, and including both color and color-airmass terms to determine the zeropoint of each image (the root-mean square of these color term fits is $\sim0.015$\,mag with respect to published SDSS magnitudes). For the PAIRITEL data, all individual frames from the same night were stacked using the Swarp\footnote{http://www.astromatic.net/software/swarp.} software. Reference frames for the subtractions were taken from the Two Micron All-Sky Survey (2MASS; Skrutskie et al. 2006) using the HOTPANTS\footnote{http://www.astro.washington.edu/users/becker/hotpants.html.} software. The subtracted images were then calibrated with respect to point sources from 2MASS. Calibration and subtraction errors were summed in quadrature.

The UVOT instrument on board {\it Swift} followed PTF11eon intensively and preliminary results from these observations were reported by Kasliwal et al. (2011) and Arcavi et al. (2011). We retrieved the level-2 UVOT data for PTF11eon from the {\it Swift} data archive. To reduce the UVOT data, the images were stacked for the individual filters on a daily basis. For the $U$, $B$, and $V$ filters, we employed the photometric calibration described by Li et al. (2006) to measure the photometry of PTF11eon in the subtracted images. For the three UV filters (UVW1, UVW2, and UVM2), the Poole et al. (2008) calibration was used. Many tests have shown that the two calibrations by Li et al. (2006) and Poole et al. (2008) are consistent with each other for bright sources.

Our photometry is corrected for Galactic extinction using the Schlegel et al. (1998) maps via NED and applying the Cardelli, Clayton, \& Mathis (1989) algorithm assuming $R_{V}=3.1$. We do not correct for host extinction, as it is found to be very small (see below). We adopt a distance modulus of 29.52 mag and present our photometry in the Vega system (except for the P48 data which are calibrated to the AB system).

\section{Analysis}

\subsection{Photometry}

The total equivalent widths of the Na~I~D$_2$ and D$_1$ lines, as measured from the HIRES spectra, was found to be $188\pm8$ m\AA\, and $107\pm6$ m\AA\, respectively. Using the correlation presented by Munari \& Zwitter (1997) for the Na~I~D$_1$ line, we constrain the local extinction value to $E(B-V)<0.05$ mag. Dust reddening is therefore negligible, and we ignore it in our analysis.

\begin{figure}
\includegraphics[width=1\textwidth]{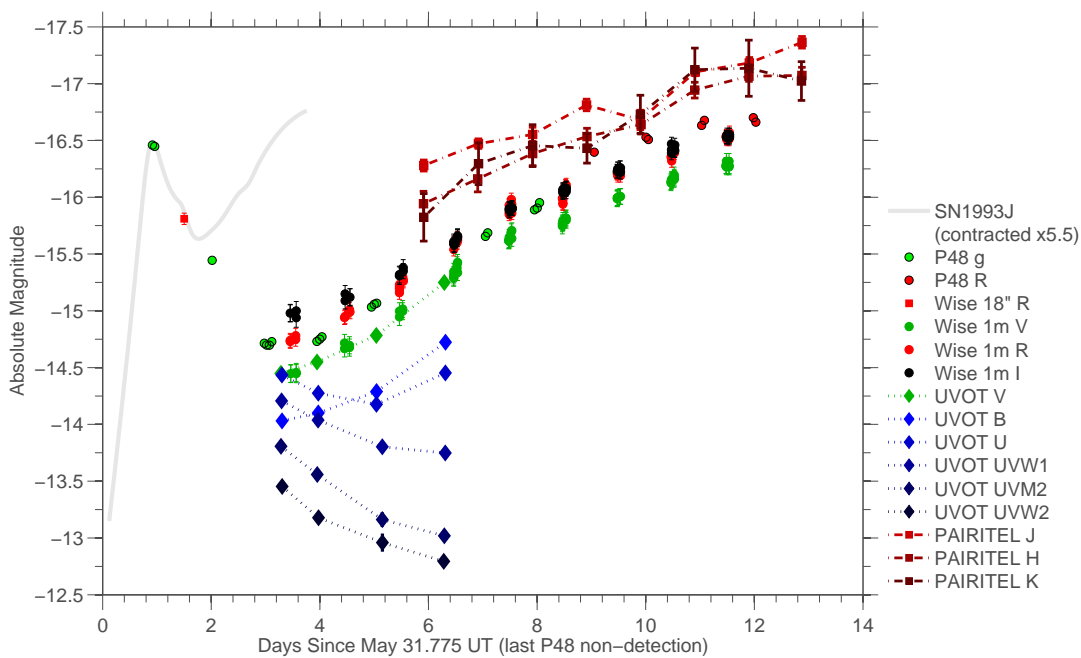}
\caption{The multi-band light curve of PTF11eon. For comparison, the early-time light curve of SN1993J from Richmond et al. (1994) is shown in grey, {\it contracted in time by a factor of 5.5}, and fainter by 0.65 mag. It is obvious that the evolution of PTF11eon at early times is much faster than that of SN1993J, and it is also possibly a little fainter (although uncertainties in distance determination to the events should be kept in mind). Such rapid evolution suggests that the progenitor of PTF11eon is much smaller than that of SN1993J. Where the error bars are not visible, they are smaller than the symbol sizes.}
\label{lc}
\end{figure}

Our photometric data are presented in Figure \ref{lc}. The initial decline in the light curve on days 1--3.5, prior to the rise to peak magnitude, can be interpreted as a detection of the shock-breakout cooling tail. Such photometric behavior is reminiscent of that seen in the Type IIb SN 1993J (Richmond et al. 1994), but PTF11eon shows a more rapid decline. This suggests that the radius of the progenitor of PTF11eon is much smaller than that of SN1993J. Indeed, using Eq. 13 from Rabinak \& Waxman (2011) with a radius of $10^{13}$\,cm, we find that two days after explosion the color temperature should be greater than 17000~K (such temperatures are indeed observed for RSG explosions such as PTF10vdl; Gal-Yam et al. 2011), while our spectrum at that time shows a much lower temperature ($7600$~K from a blackbody fit as well as the presence of low-ionization elements; see Fig. \ref{spec}). In addition, the sharp decline in the $g$ band light curve requires a much lower temperature even before day two. We note that the effective temperature, which is $\sim20$\,\% lower than the color temperature, is still higher than the recombination temperature of H, indicating that the assumption of ionized ejecta made by the model is consistent.

It is therefore highly unlikely that the radius of the progenitor of PTF11eon was of order $10^{13}$\,cm, as claimed by Maund et al. (2011) and Prieto \& Hornoch (2011). Therefore, the object identified on pre-explosion HST images (Maund et al. 2011, Van Dyk et al. 2011) is probably not the solo progenitor of PTF11eon. Rather, the progenitor could be in a binary system (Van Dyk et al. 2011) or a member of a compact cluster. A relatively faint Wolf-Rayet star could satisfy a radius as small as $10^{11}$\,cm, while possibly remaining below the magnitude limits imposed by the {\it HST} data.

Chevalier \& Soderberg (2010) estimate a radius of $\sim10^{11}$\,cm for the progenitor of SN 2008ax, which shows similar spectral properties to those of PTF11eon (Fig. \ref{spec_comp}). Pastorello et al. (2008) constrain the length of the shock-breakout cooling phase of SN2008ax to be less than a few hours, implying that the radius of the progenitor of PTF11eon could be larger than that of SN2008ax (yet still much smaller than that of SN1993J). We leave detailed modeling of the shock-breakout cooling phase to a future report.

\subsection{Spectroscopy}

\begin{figure}
\includegraphics[width=1\textwidth]{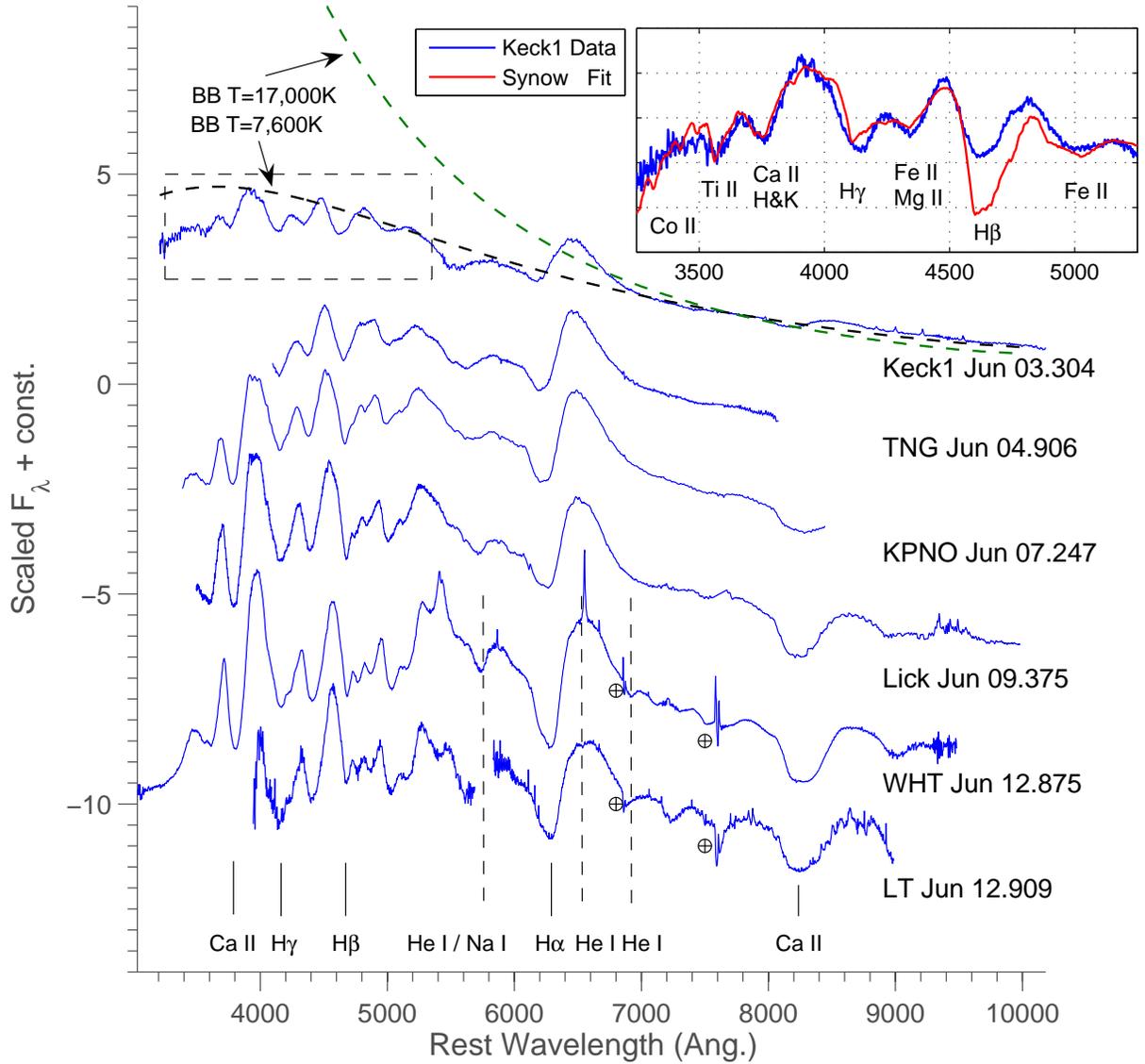}
\caption{Spectra of PTF11eon. The first spectrum, taken roughly two days after explosion, shows highly developed low-ionization features and a blackbody temperature of $\sim 7600$\,K (dashed black line). It is highly inconsistent with a blackbody temperature of $17000$\,K (dashed green line) implied by a progenitor star with radius $10^{13}$\,cm. The last two spectra, taken on the same night, show possible He I lines emerging (the $6678$\,\AA\, line is obscured by the narrow H$\alpha$ feature from the underlying host galaxy in the WHT spectrum, but is present in the LT IFU spectrum which shows much less host contamination). The H lines are plotted at $13000$\,km\,s$^{-1}$, the He I lines at $6500$\,km\,s$^{-1}$, and the Ca~II lines at $12000$\,km\,s$^{-1}$. Telluric absorption features are marked. Inset: A SYNOW fit to the blue side of the Keck spectrum.}
\label{spec}
\end{figure}

\begin{figure}
\includegraphics[width=1\textwidth]{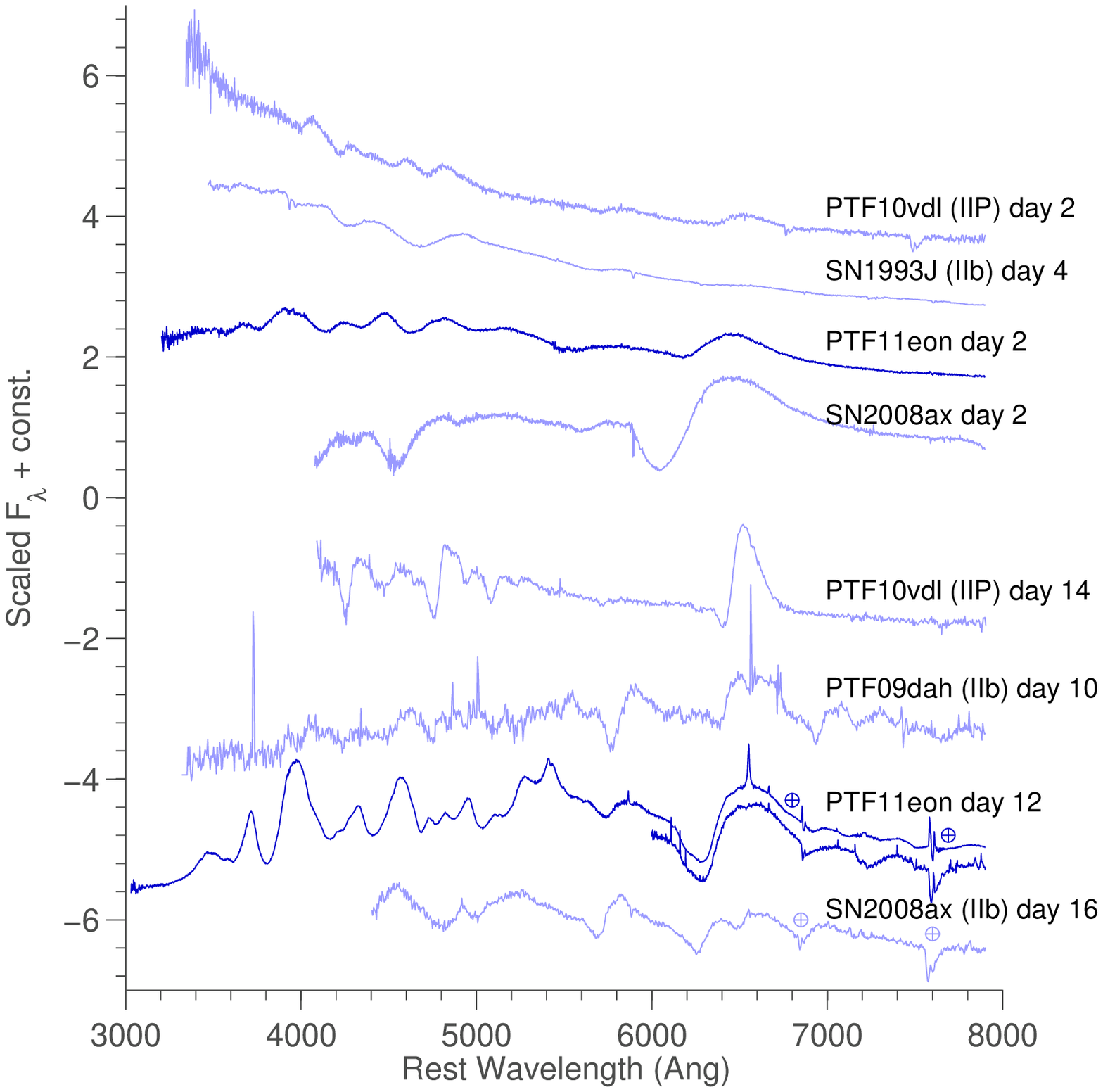}
\caption{Comparison of the spectra of PTF11eon at roughly 2 and 12 days with PTF10vdl (a Type IIP SN; Gal-Yam et al. 2011), SN1993J (Type eIIb; spectrum obtained by the Asiago SN Group and retrieved via SUSPECT), SN2008ax (Type cIIb; early spectrum from Chornock et al. 2010, late spectrum from Pastorello et al. 2008) and PTF09dah (a Type IIb very similar to SN 1993J; Arcavi et al., in prep.). In the early spectra (top) PTF11eon is not as blue as PTF10vdl and SN1993J and shows more developed features (including neutral H lines) compared to these SNe, as expected from a smaller progenitor. SN2008ax shows similar properties, however at higher velocities, consistent with an even smaller radius. At later times (bottom) PTF11eon does not yet show prominent He features characteristic of Type IIb events such as PTF11dah, however these lines developed a few days later (Marion et al. 2011).}
\label{spec_comp}
\end{figure}

We used SYNOW (Jeffery \& Branch 1990) to infer elements in the spectra of PTF11eon. The spectra show prominent H features as well as Ca II and Fe-peak elements characteristic of Type II SNe (Fig. \ref{spec}). We see these features at velocities as high as $\sim17000$\,km\,s$^{-1}$ initially and at an earlier stage compared to other young Type II SNe, which tend to show a blue continuum with low-contrast H$\alpha$ emission a few days after explosion (Fig. \ref{spec_comp}). Using a blackbody fit to the June 3 spectrum, we find a temperature of $\sim 7600$\,K. We note that, unlike SN1993J, no prominent He features are seen in the spectra of PTF11eon, even out to day 10 after explosion. However, hints of He absorption might be present in the spectra of June 12 (Fig. \ref{spec}). The similarity of the PTF11eon spectrum on day 12 to that of the Type IIb SN 2008ax on day 16 (Fig. \ref{spec_comp}) suggests that these features may develop to become clear He signatures (as confirmed by Marion et al. 2011). We conclude that PTF11eon is likely a Type IIb supernova from a compact progenitor (termed cIIb by Chevalier \& Soderberg 2010).

\section{Concluding Remarks}

We have presented our discovery of PTF11eon / SN2011dh, a supernova in M51, as well as results from our early follow-up effort. This event displays similarities to the Type IIb SNe 1993J and 2008ax, but also important differences. The rapid early-time evolution of the light curve compared to that of SN 1993J and the relatively low temperature implied by our first spectrum indicate that PTF11eon resulted from the explosion of a fairly compact star, not consistent with the object identified by Maund et al. (2011) as a single star. Rather, as suggested by Van Dyk et al. (2011), the {\it progenitor system} is possibly composed of a binary system of which the progenitor of PTF11eon is a member. PTF11eon demonstrates the importance of real-time detections and of rapid and continuous follow-up observations, as crucial tools for constraining both explosion and progenitor properties via shock breakout analysis (Soderberg et al. 2008, Schawinski et al. 2008, Gezari et al. 2008, Ofek et al. 2010). Obtaining and analyzing a statistical sample of such early supernovae will serve to probe the progenitor stars of a much larger sample, beyond the $\sim10$ Mpc limit for direct progenitor detections, placing strong and novel constraints on the last phases of massive stellar evolution. PTF continues to collect data for this purpose.

\section{Acknowledgments}

The Weizmann Institute PTF partnership is supported by the Israeli Science Foundation via grants to A.G. Collaborative work between A.G. and S.R.K. is supported by the US-Israel Binational Science Foundation. A.G. further acknowledges support from the EU FP7 Marie Curie program via an IRG fellowship, the Benoziyo Center for Astrophysics and a Minerva grant. P.E.N. is supported by the US Department of Energy Scientific Discovery through Advanced Computing program under contract DE-FG02-06ER06-04. M.S. acknowledges support from the Royal Society; M.S. and A.G. are also grateful for a Weizmann-UK Making Connections grant. A.V.F.'s supernova group at U.C. Berkeley acknowledges generous support from Gary and Cynthia Bengier, the Richard and Rhoda Goldman Fund, US National Science Foundation grant AST-0908886, and the TABASGO Foundation. J.S.B. was partially supported by a SciDAC grant from the U.S. Department of Energy and a grant from the National Science Foundation (award 0941742). P.M., E.P.,  and E.S.W. acknowledge financial support from INAF through PRIN INAF 2009.

The C18 telescope and most of its equipment were acquired with a grant from the Israel Space Agency (ISA) to operate a Near-Earth Asteroid Knowledge Center at Tel Aviv University.

LAIWO, a wide-angle camera operating on the 1-m telescope at the Wise Observatory, Israel, was built at the Max Planck Institute for Astronomy (MPIA) in Heidelberg, Germany, with financial support from the MPIA, and grants from the German Israeli Science Foundation for Research and Development, and from the Israel Science Foundation.

The WHT is operated on the island of La Palma by the Isaac Newton Group in the Spanish Observatorio del Roque de los Muchachos of the Instituto de Astrofísica de Canarias.

The National Energy Research Scientific Computing Center, which is supported by the Office of Science of the U.S. Department of Energy under Contract No. DE-AC02-05CH11231, provided staff, computational resources, and data storage for this project.

\end{document}